\documentclass{sigchi}

\CopyrightYear{2018}
\doi{}
\isbn{}
\conferenceinfo{}{}
\acmPrice{}
\usepackage{float}

\usepackage{balance}       
\usepackage{graphics}      
\usepackage[T1]{fontenc}   
\usepackage{txfonts}
\usepackage{mathptmx}
\usepackage[pdflang={en-US},pdftex]{hyperref}
\usepackage{color}
\usepackage{booktabs}
\usepackage{textcomp}
\usepackage{array}
\usepackage{tikz}
\usetikzlibrary{shapes.geometric, arrows}
\usepackage{multirow}
\usepackage{hhline,booktabs,amsmath}
\usepackage{makecell}
\usepackage{microtype}        
\usepackage{ccicons}          
\usepackage{graphicx}
\usepackage{todonotes}

\def\plaintitle{SIGCHI Conference Proceedings Format}

\def\emptyauthor{}
\def\plainkeywords{}

\makeatletter
\def\url@leostyle{%
  \@ifundefined{selectfont}{
    \def\UrlFont{\sf}
  }{
    \def\UrlFont{\small\bf\ttfamily}
  }}
\makeatother
\def\pprw{8.5in}
\def\pprh{11in}

\definecolor{linkColor}{RGB}{6,125,233}
\hypersetup{%
  pdftitle={\plaintitle},
  pdfauthor={\emptyauthor},
  pdfkeywords={\plainkeywords},
  pdfdisplaydoctitle=true, 
  bookmarksnumbered,
  pdfstartview={FitH},
  colorlinks,
  citecolor=black,
  filecolor=black,
  linkcolor=black,
  urlcolor=linkColor,
  breaklinks=true,
  hypertexnames=false
}

\begin{document}

\title{A clustering approach to infer Wikipedia contributors' profile}

\numberofauthors{3}
\author{%
  \alignauthor{Shubham Krishna\\
    \affaddr{IIT(ISM)}\\
    \affaddr{Dhanbad, India}\\
    \email{shubhamkrishna@am.ism.ac.in}}\\
  \alignauthor{Romain Billot\\
    \affaddr{IMT Atlantique-UBL \& Lab-STICC}\\
    \affaddr{Brest, France}\\
    \email{Romain.Billot@imt-atlantique.fr}}\\
  \alignauthor{Nicolas Jullien\\
    \affaddr{IMT Atlantique-UBL, LEGO-M@rsouin,}\\
    \affaddr{Brest, France}\\
    \email{Nicolas.Jullien@imt-atlantique.fr}}\\
}

\maketitle

\begin{abstract}
In online communities, recent studies have strongly improved our knowledge about the different types or profiles of contributors, from casual to very involved ones, through focused people. However they do so by using very complex methodologies (qualitative-quantitative mix, with a high workload to manually codify/characterize the edits), making their replication for the practitioners limited. These studies are on the English Wikipedia only. The objective of this paper is to highlight different profiles of contributors with clustering techniques. The originality is to show how using only the edits, and their distribution over time, allows to build these contributors profiles with a good accuracy and stability amongst languages. The methodology is validated with both Romanian and Danish wikis. The highlighted profiles are identifiable early in the history of involvement, suggesting that light monitoring of newcomers may be sufficient to adapt the interaction with them and increase the retention rate. 

\end{abstract}
\category{Social and professional topics}{ Management of computing and information systems}{Project and people management- Project staffing}
\category{Human-centered computing}{ Collaborative and social computing}{   Collaborative content creation}
\category{Human-centered computing}{ Empirical studies in collaborative and social computing}{}
\category{Human-centered computing}{                                                  Collaborative and social computing systems and tools}{ Wikis}


\maketitle
\section{Introduction}

In the open online communities, such as Free, Libre, and Open Source (FLOSS)\cite{ArafatRiehle09}, it has been shown that the number of articles and edits per author follows a power law \cite{Voss05,JohnsonFarajKudaravalli14}, like in scientific publication \cite{Maillartetal08}. Even for Wikipedia, which claims that it is 'the encyclopedia that everybody can edit', this repartition exists \cite{Kitturetal07b,Ortegaetal09,Zhangetal10}, and 'the top 10\% of editors by number of edits contributed 86\% of the PWVs {[}persistent word views{]}, and top 0.1\% contributed 44\% - nearly half! The domination of these very top contributors is increasing over time.' \cite[p. 5]{Priedhorskyetal07}. 
This apparent paradox is easy to understand: contributing to an online community is not only about having something to say but more
and more about knowing how to say it \cite{FordGeiger2012}.
And, as Wikipedia has become bigger, the editing tasks have increased in complexity (see \cite{FongBiuk-Aghai10} for a proposition of classification in terms of semantic complexity of these various type of edits), and have increased also the proportion of non-editing tasks. In other words, participants' types of activity have multiplied. 
Beyond the writing, which can be seen as the emerged part of the iceberg, but also the most important part, for an encyclopedia, are the actions leading to the writing (coordination tasks, discussions on the topic of the project, etc.), but also the actions of maintaining the existent, which take a growing importance with the maturation of the articles \cite{KaneJohnsonMajchrzak14}.

One consequence is that over time, the amount of effort needed to add new content increases, since new edits are more likely to be rejected, making the work less rewarding \cite{RansbothamKane11,AaltonenSeitler16}.
This may also explain, for a part, the contributor turnover \cite{Farajetal16,KaneJohnsonMajchrzak14}: once a project is finished, or at least mature, some people, those interested in content addition, drop. As a consequence, there is a constant need for these projects to recruit new contributors, and to turn them into 'big' contributors, to guarantee the survival of the project in the long run.

There are a lot of experiments to slowly engage people into contribution (from simple edits to more complex tasks), based on the concept of legitimate peripheral participation \cite{LaveWenger91,Wengeretal02}. For example, and still on Wikipedia, some experiments show that readers or contributors can be asked to perform small tasks, that they do, and then keep participating \cite{HalfakerKeyesTaraborelli13}. Acknowledging the newcomers contribution with moral reward ('banstars') increases their investment and their retention, at least over the first year \cite{Gallus15}.
But it may not be a very sustainable activity, as those who respond the most to these kinds of initiative seem to be those who are already willing to participate \cite{Narayanetal17}.
And, statistically speaking, big contributors seem to have been so from the beginning, and if there is a path to contribution, it concerns more the learning of the rules than the level of contribution \cite{PancieraHalfakerTerveen09,DejeanJullien15}.

In a nutshell, beyond this statistical information of a majority of big contributors from the beginning, what these experiments seem to indicate is the existence of different profile of contributors regarding their involvement. And one may want to know better about these different profiles and if they are in the same proportion amongst all the projects.
This is important for the managers of such projects. It would allow them to  better adapt their response to newcomers contributions, and to improve their retention rate.
Recent studies \cite{Weietal15,Yangetal16, Arazyetal16} have strongly improved our knowledge of the different types or profiles of contributors, from casual to very involved ones, through focused people. However they do so by using very complex methodologies (qualitative-quantitative mix, with a high workload to manually codify/characterize the edits), making their replication for the practitioners limited. These studies are on the English Wikipedia only. The objective of this paper is to highlight different profiles of contributors with clustering techniques. The originality is to show how using only the edits, and their distribution over time, allows to build these contributors profiles with a good accuracy and stability amongst languages. These profiles are identifiable early in the history of involvement, suggesting that light monitoring of newcomers may be sufficient to adapt the interaction with them and increase the retention rate.


The paper continues as follows: the next section reviews our theoretical background to develop our hypotheses regarding the profiles of the contributors, and the good balance between the simplicity of the variables and the accuracy of the results. Then, we describe our data collection strategy (choice of Wikipedia, data and variables), before presenting the methodology and the results. Finally, we discuss our findings and highlight their implications for both theory and practice, before concluding.

\section{Research hypotheses: Contributor's Behaviors and Roles Detection}

It has been no more a matter of debate that regular contributors vary in the tasks they perform, leading to various 'career' within the projects.  For instance, \cite{OkoliOh07}, looking at English Wikipedia contributors, showed that people having lots of participation in various
articles (and thus collaborating with a lot of people, but not in a sustained manner, something they assimilate to 'weak links', in a Granovetter's perspective \cite{Granovetter85}) are more likely to become administrators (to have administrative rights) than those more focused on a sub-set of articles and talking repeatedly with a small subset of people (and then developing strong(er) links). This leads \cite{Zhuetal11}, relying on \cite{Bryantetal05}'s study, to propose two main careers for the Wikipedia contributors, coherent with \cite{OkoliOh07}'s findings: from non-administrators to administrators and from non-members to Wiki-project regular members to Wiki-project core members (Figure 1, page 3433). On that aspect, \cite{AntinCheshireNov12} confirmed that people involved from the beginning in more diverse revision activities are more likely to take administrative responsibilities.

Qualitative research has refined our understanding of people's interest and focus: in their in-depth analysis of one English Wikipedia article (autism) \cite{KaneJohnsonMajchrzak14} showed that in the lifetime of an article, different tasks were required (contend edition, article structuring, knowledge stability protection), requiring different skills and centers of interest, and consequently endorsed by different persons, with different level of edits.

\cite{Huvila10}, using a ground theory approach via an online open-question survey to contributors, proposed a classification in five types for the contributors, according to their activities and to the way they find their information (from in depth research to personal/professional area of expertise, through browsing the net). These profiles illustrate how diverse even contributing knowledge can be, between the topics, but also between the sources of information they rely on, but the contributing profiles remain: some people focus on an area of expertise, other contribute a lot on a lot of subjects, others are more casual, etc.

Informed by these findings, several authors proposed qualitative techniques to retrieve and quantify the different roles the qualitative research has identified. 
Being able to do quantitative identification makes its automatizing possible, and can then decrease the supervision burden, in addition to increase its accuracy and its rapidity.
However, as for article quality identification\footnote{It is well known, since \cite{WilkinsonHuberman07} that there is a strong correlation between the number of edits and the probability for an (English) Wikipedia article to be of best quality. Nevertheless, as detailed by \cite{DangIgnat17}, if one wants to refine this finding, more costly methods are needed in terms of data collection and analytic techniques}, there is a debate between the simplicity and the accuracy of the methods used.
What is directly observable, in most of the open, online projects, is the number of contributions (edits, commits) over time. What is less accessible, requiring more data preparation, and most of the time, allowing only ex-post analyses, is the content, and the quality, of such contributions. 
\cite{Yangetal16}, in a defense of the second trail of research, summarized this trade-off, as follows: 
'While classification based on edit histories can be constructed for most active editors, current approaches focus on simple edit counts and access privileges fail to provide a finer grained description of the work actually performed in an edit'.

And it is to be acknowledged that, as far as the English Wikipedia is concerned, research has made tremendous progress. Via a mix of non supervised and supervised techniques \cite{Yangetal16, Arazyetal16,Weietal15}, scholars identified and characterized the edits, and then constructed editor roles based on their characterized edits.
Looking at the English Wikipedia, \cite{Yangetal16} proposed a two-steps methodology. First, to enrich the description of the edits, they used a multi-class classifier to assign edit types for edits, based on a training data set, called "the Annotated Edit Category Corpus" they annotated themselves. Then they applied a LDA graphical model, in order, in a second step, to identify editors' repeating patterns of activity, and to cluster the editors according to their editing behaviors. Afterwards, the authors try to link these behaviors to the improvement of articles quality.
\cite{Arazyetal16}, clustered, on a stratified sub-set of a thousand English Wikipedia articles, the contributors according to their edits, edits classified using supervised learning techniques. They confirmed and refined the above qualitative results. They also showed, in \cite{Arazyetal17}, that some people can take different role over time, when others stick to the same behavior in the various articles they contribute to. 

In citizen science, \cite{Jacksonetal16} used a similar approach to study the newcomers' activities (contributing sessions) and clustered their behavior in a Zooniverse project (a citizen science contributive project), Planet Hunters, "an online astronomy citizen science project, in which astronomers seek the help of volunteers to filter data collected from the Kepler space telescope". Based on a mix qualitative-quantitative methods, they first observe and interviewed participant regarding their contributing behaviors, in order to define the tasks to be observed to define a contributing pattern. Then, they aggregated page view data and server logs containing annotations and comments of each participant, and regrouped data by activity by 'session' (a session was defined as "a submission of an annotation where no more than 30 minutes exists between the current and next annotation"). They clustered the sessions based on counts of dimensions (e.g., number of contributions to /object, /discussion, annotations), using a k-means clustering algorithm to defined types of sessions, and, finally, they described the people by their history of participation (the types of sessions they did). Interestingly, the type of sessions and the contributor profile they found are very similar to those found in Wikipedia\footnote{Their principal finding are 1) that many newcomers remained in a single session type (so they can be detected quite early in their participation journey); 2) that the contributor patterns can be regrouped in three types: Casual Workers, Community Workers, and Focused Workers.}. 

Even if these studies can be extended toward other case studies than English speaking projects, it is not sure they could go farther in terms of precision in the description of the different profiles, neither that the people involved in those projects will be aiming at investing their time to manually create the dataset of coded contributions these methods require.
We argue that there is still some work to do in the detection of these profiles, especially amongst newcomers, but more on the simplification of the detection methods rather that in their over-sophistication. 
What our discussion shows is that practitioners and researchers have the too extremities of the story: the newcomers seem to engage themselves in a contributing profile very early in their contributing history; they converge toward different contributing profiles.
But how much data do we need to connect the dots, and how early is it possible to do so? There are strong managerial reasons for advocating for detection as early as possible, with not too much apparatus.

For being able to quickly respond to contributors/newcomers, community managers need not ex-post data analysis (which are very good to describe the behaviors), but tools to identify people along the way, to adapt the interactions as soon as possible.
If the development of complex, artificial intelligent tools is very high for the 'big' Wikipedias, it is slower for the smaller ones, and its tuning is based on the cultural and organizational principals of those big Wikipedias, and especially of the English one\footnote{Such as the ORES project, on detecting the quality of the edits, which is quite developed over the different Wikipedias for detecting the quality of an edit, but not very much at article level \url{https://www.mediawiki.org/wiki/ORES}.}.



This calls for a temporal description of the contributors, with the minimum of data extraction, or qualification (and possibly the data which are already available for consultation). As stressed by all these studies, the minimal information is the contribution (commit, edit), and all the studies cited are based on the analysis of the edits (sometimes the enriched edits).
As a consequence, we wonder, in this article, if observing only the edit behavior over time, it is possible to distinguish different profiles, and, if so, to link these profiles with the ones detailed by more complex methods. We put ourselves in the path of \cite{Welseretal11} and of other sociological studies \cite{GusevaRona-Tas01}, wondering if it is possible to find 'structural signatures of social attributes of actors'.

\section{Research methodology}
\begin{figure}[!h]
\begin{center}
\includegraphics[width=12cm]{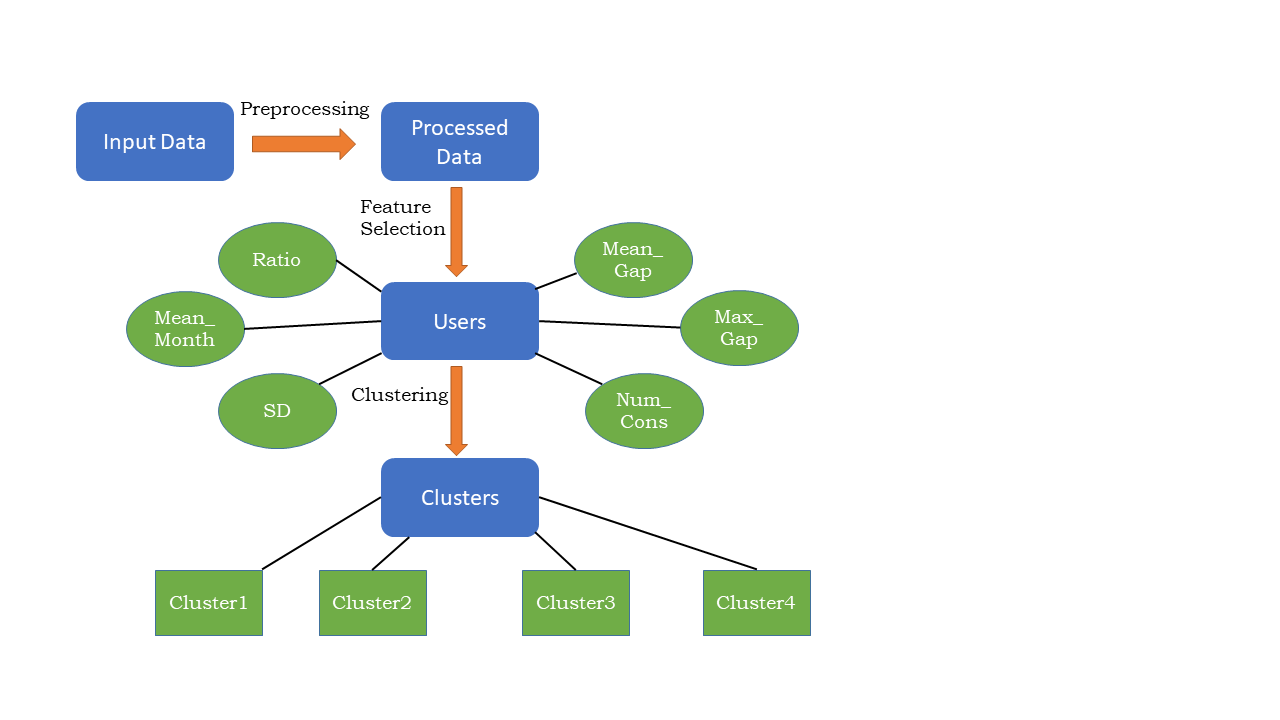}
\end{center}
\caption{Methodological flowchart.}
\label{flowchart}
\end{figure}

This section will describe the different steps that compose our research methodology, from raw data to the interpretation of the 4 clusters in terms of contributors activity and roles. Figure \ref{flowchart} gives the global picture of the methodology
detailed in the next subsections.\footnote{Figure inspired by \cite{Jacksonetal16}.}
\subsection{Data collection strategy}
One of the most useful thing about Wikipedia is that many data are publicly available for downloading and analyzing. This includes information about Wikipedia content, discussion pages, contributor pages, editing activity (who, what and when), administrative tasks (reviewing content, blocking users, deleting or moving pages), and many other details\footnote{It has to be understood that when speaking of "user" page, Wikipedia means the users of the wiki, or, more simply the contributors. The simple readers are called readers. In this article we use the terms contributor and user indifferently}. There are many different ways of retrieving data from Wikipedia such as web-crawlers, using available APIs and etc. 
We used database dump files which are publicly available for every language and can be downloaded from Wikimedia Downloads center. An important advantage of retrieving information from these dump files is that researchers have complete flexibility as for the type and format of the information they want to obtain. These dump files are usually available in XML and SQL formats. An important remark about these dump files is that every new file includes again all data already stored in prior versions plus the new changes performed in the system since the last dump process, and excluding all information and meta-data pertaining pages that have been deleted in that interval.



In our research, we studied Danish and Romanian Wikipedia to show how our methodology can be implemented on mid-size language projects. The required data for our analysis was present in the "pages-meta-history" dump file which was completed on 1st January, 2018. This dump file contains data about complete Wikipedia text and meta-data for every change in the Wikipedia from the launch of that Wikipedia till December,2017. After getting  the dump file, we used WikiDAT\footnote{\url{http://glimmerphoenix.github.io/WikiDAT/}} for extraction of data from the dumps. Wikipedia Data Analysis Toolkit abbreviated to WikiDAT is a tool that automates the extraction and preparation of Wikipedia data into 5 different tables of MySQL database (page, people, revision, revision hash, logging). WikiDAT uses Python and MySQL database and was developed with the motive to create an extensible toolkit for Wikipedia Data Analysis.

\subsection{Construction of the variables}
In the field of pattern recognition, it is very important to have features that are informative, discriminative and should explain the variability present in the data. As a primary data filtering step, the study has been limited to those contributors who have contributed more than 100 edits (irrespective of whether the edits made by them were minor or major) on the respective Wikipedias\footnote{The definition of what a contributor is still a matter of debate. \cite{PancieraHalfakerTerveen09}, studying the English Wikipedia, defined a "Wikipedians", or a regular, really involved contributors, as people having made at least 250 in their lifetime. We chose a smaller Figure because we wanted to capture the behavior of the not so involved, in a nutshell, all those who have been active for several months. And an "editor", for the Wikimedia Foundation, is somebody who have contributed 5 edits or more in a month. We also wanted to have a big enough number of contributors.}. We removed those contributors who were either robots, contributed only in a single month and contributed anonymously. There were such 171  contributors in the Romanian Wikipedia and 274 contributors in the Danish Wikipedia. 
As said, our goal was to use simple activity measures based only on the edits and their distribution over time. With respect to the state of the art, contributors are likely to be grouped in terms of volume, intensity (focus) or duration of the activity. Starting with a "brainstorming" list of 12 initial features, a short list of 6 features was obtained after studying the correlation matrix. The features that were dropped were the number of edits/contributions made, the number of days user has been on wikipedia, the minimum and the median gap between two consecutive posts, the median number of edits/contributions made during different months and the number of different months a user has contributed in. Indeed, redundancy has been dropped by removing some heavily correlated features. The final features used for the statistical analysis are described in Table \ref{tab:Variable-description}.




\begin{table}[!h]
\caption{\label{tab:Variable-description}Description of the variables}
\begin{tabular}{|c|m{6 cm}|}
\hline 
Variable & Description\tabularnewline
\hline 
\hline 
Ratio & The ratio between the number of edits and the number of days a contributor has been on Wikipedia from the very first edit \tabularnewline
\hline 
Mean\textunderscore gap & The average gap between two consecutive posts measured in months\tabularnewline
\hline 
Max\textunderscore gap & The maximum gap between any two consecutive posts measured in months. \tabularnewline
\hline 
Num\textunderscore cons & The number of pairs of consecutive months with contributions \tabularnewline
\hline 
Mean\textunderscore Month & Per month average edits made \tabularnewline
\hline 

SD &  Standard Deviation among the month average edits value \tabularnewline
\hline 
\end{tabular}
\end{table}

Ratio is a measure of how massively the contributors have contributed during their entire period of contribution and it incorporates with it the relationship between the number of edits and the number of days. Mean\textunderscore Month provides information about average number of edits made in a month and SD tells us about the variations in the contributions made during these months. Collectively, we can say that the features Ratio, Mean\textunderscore Month and SD is an evaluation about the quantity and deviation of the contributions made by the contributors. Mean\textunderscore Gap is a measure that describes the average time gap between two consecutive posts and Max\textunderscore gap is a measure about the longest period of inactivity between two successive posts. Both features give us information about how often the contributors get active and how long they can quit the community before coming back. The feature Num\textunderscore cons tells us about how many times the contributors have contributed successively for two consecutive months. For example, if a contributor made edits in January 2011 and February 2011, the count is increased by 1. In other words, Num\textunderscore cons is a measure of regularity of contributors edits over time.

\subsection{Statistical methods}
Clustering techniques were used to group contributors in similar clusters highlighting various pattern in terms of activity and roles. In order to design robust conclusions, the Romanian Wikipedia was used to calibrate the methods and come up with a first groups interpretation. Then, the Danish Wikipedia was used as a validation dataset to check the group correspondence across different datasets. A contribution of this article is to provide this double checking in terms of cluster validation. Regarding the methods, a two-stage cluster analysis was performed:
\begin{enumerate}
\item A hierarchical clustering was done based on the features described in the Table \ref{tab:Variable-description} with the \textit{hclust} function of the R platform for statistical computing \cite{RRR}. The metric used was the Ward distance, adapted to quantitative features \cite{duda2012pattern}. The resulting dendrogram can suggest a first trend about the optimal number of clusters, in terms of loss of inertia.

\item Partitioning algorithms were used as alternative clustering methods in order to select the final typology.  In our research, the contributors were clustered  using a k-medoids clustering algorithm called PAM (Partitioning Around Medoids), from the R package \textit{cluster}. The PAM algorithm is based on the search for $k$ representative objects or medoids among the observations of the data set. It is known to be more robust than the k-means algorithm, especially with respect to the initialization \cite{kaufman2009finding}.

In this work, different typologies have been formed for $k$ ranging in the interval selected in step 1. Results of the PAM algorithms were consistent with those obtained from step 1. Then, the optimal number of cluster has been selected with cluster validation technique such as the silhouette index \cite{halkidi2002cluster} which models how well contributors are clustered into their groups (intra vs. inter cluster inertia). 
\end{enumerate}

To assist the interpretation of the resulting clusters, Principal Component Analysis (PCA) has been carried out in order to project the data onto a small number of dimensions that are combinations of the initial variables \cite{saporta2006probabilites}. In this article, three dimensions were enough to explain almost $90\%$ of the data variability. In addition to PCA, ANOVA analysis and Tukey statistical tests have helped to determine the significant variables within each cluster.  All together, these methods ensure a full and robust interpretation of clustering results.

\section{Results}
\begin{figure}[!h]
\begin{center}
\includegraphics[width=10cm]{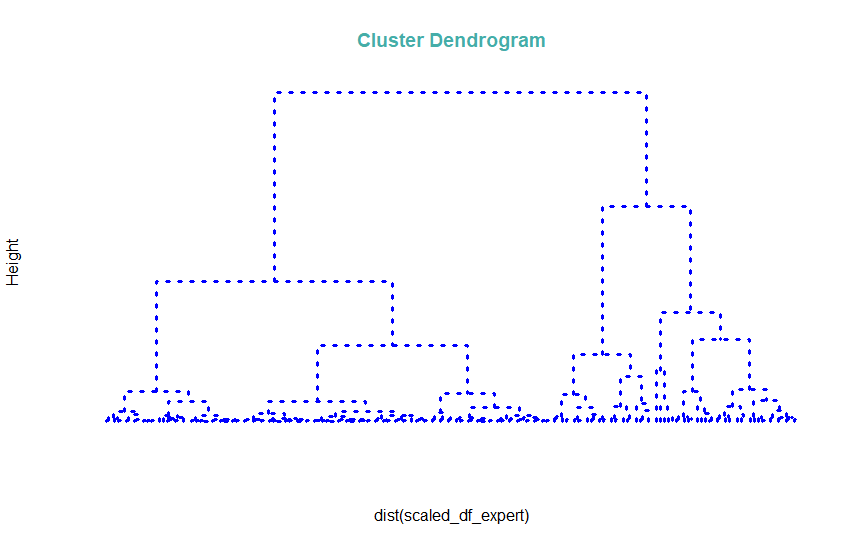}
\end{center}
\caption{Cluster dendrogram of the Romanian Wikipedia}
\label{den}
\end{figure}

\begin{figure}[!h]
\begin{center}
\includegraphics[width=8cm]{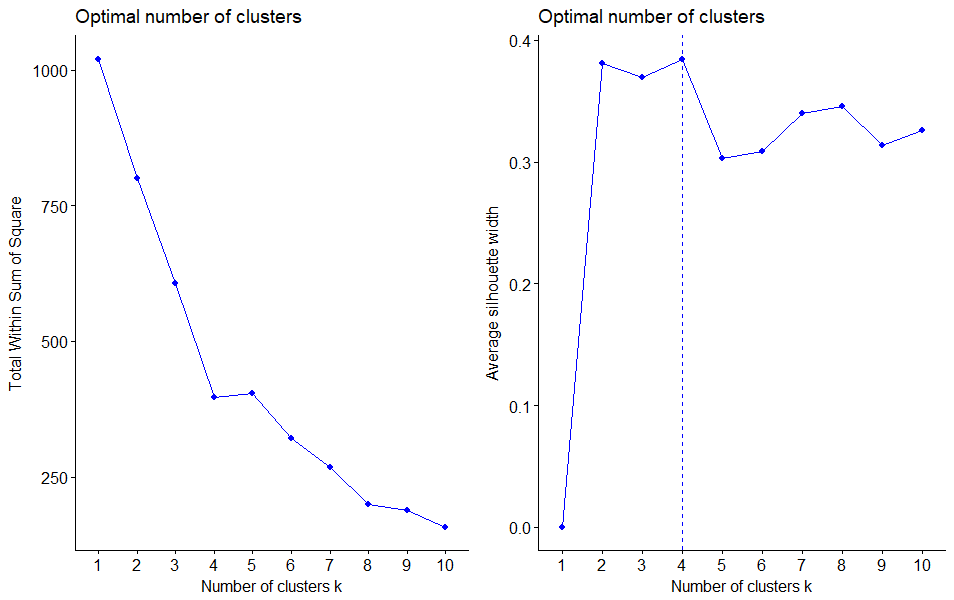}
\end{center}
\caption{Cluster Validation Plots}
\label{validity}
\end{figure}

\begin{figure*}[!h]
\begin{center}
\includegraphics[width=\textwidth]{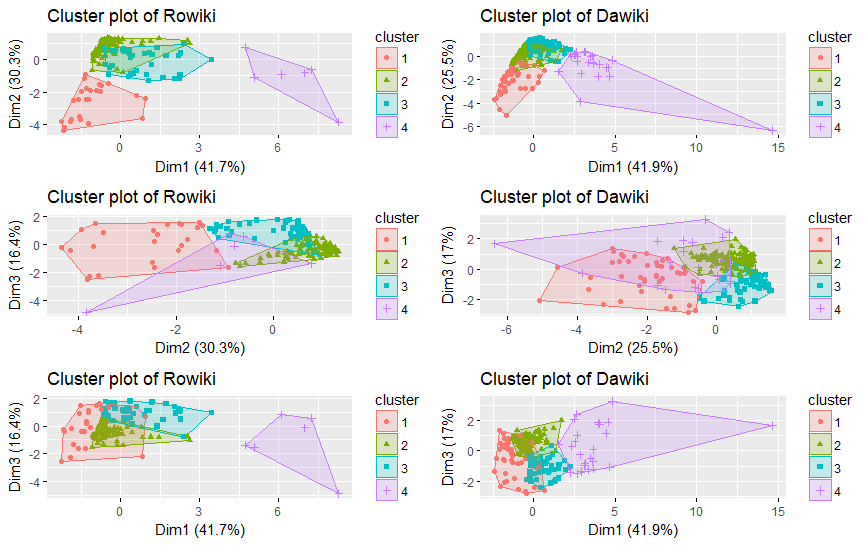}
\end{center}
\caption{PCA Analysis with projection of the four clusters}
\label{PCA}
\end{figure*}

Hierarchical clustering gives a first visualization of the data structure (Figure \ref{den}). As for the optimal number of clusters $k$, the dendrogram suggests an interval between 2 and 10 clusters that could be investigated by further evaluation.
As mentioned in the previous section, the evaluation has been made with cluster validity indexes such as the average silhouette width and the total within sum of squares. Figure \ref{validity} depicts the evaluation results in four clusters of contributor's contribution behavior in Romanian Wikipedia, this number being validated afterward with the Danish Wikipedia.
 One cluster in our analysis contains the least number of contributors in both the cases. The distribution of the contributors in the clusters for both the Wikipedias are in Table \ref{Table:Size_of_Clusters}. 
\begin{table}[h] 
\centering
\caption{Size of Clusters}
\label{Table:Size_of_Clusters}
\begin{tabular}{l c c c c }
\hline
Wikipedia &  Cluster 1 & Cluster 2 & Cluster 3 & Cluster 4  \\
\hline 
Romanian& 25 & 92 & 48 &  6 \\
Danish& 45 & 144 & 61 &  24\\
\hline
\end{tabular}
\end{table}
With respect to clusters interpretation, a PCA with three principal components explains almost 90\% of the total dataset variance. Figure \ref{PCA} depicts the projection of the labeled contributors onto these three first dimensions. Analyzing the loadings for both wikis, it turns out that the first dimension (PC1) is correlated with the volume of the activity (ratio, mean number of edits) with a relative intra-cluster variability. Dimension 2 (PC2) relates to the periods of inactivity (the gaps). The correlation is negative in Figure \ref{PCA}. Dimension 3 (PC3) mainly refers to the variable Num\textunderscore cons, it relates to the notion of regularity. Please note that due to computational details, this correlation is positive for the Romanian Wikipedia and negative for the Danish Wikipedia. 

The interpretation enables the extraction of the following contributors profiles:
\begin{itemize}
\item Cluster 1: contributors "on a mission". After joining the community or editing first articles, these contributors left Wikipedia for long periods. When they did contribute, they did it with a high, but short-term, activity.
\item Cluster 2: basic, or 'casual' contributors. This group contains basic characteristics with no significant markers, beside the fact that their activity is neither particularly intense, nor particularly focused (in terms of period, or of number of articles addressed).
\item Cluster 3: regular contributors. The activity is above the average (even if not that much) and the most regular among all groups, especially in terms of the number of consecutive months of presence. 
\item Cluster 4: top contributors. The contributors present a huge activity ratio, they are those core, or very active contributors found in other research articles. Nevertheless, this cluster contains higher variability than others.
\end{itemize}

These interpretations are confirmed by unidimensional boxplots distributions (Figure \ref{dawiki_boxplots} and Figure \ref{rowiki_boxplots}, in Appendix).

Generally, boxplots give a fine picture of the features distributions within each cluster, with a focus on the intra cluster variability. An illustrative variable was added to the analysis: the number of different articles a contributor has contributed to. This external feature confirms the analysis above.


\section{Discussion}

\subsection{Simple methods = solid conclusions}
Our goal was to evaluate if, with simple measures of contributing activities over time, it was possible to detect the different profiles of contributors with data reduction techniques. At least on the Wikipedia example, we have been able to detect the focused workers (cluster 1), the casual workers (cluster 2), and the regular workers (clusters 3 and 4), and even to discriminate between those the very involved (the top, or very top contributors \cite{Priedhorskyetal07}). As far as the objective is to identify contributors profile, our article shows that following the edits is quite enough. The number of articles involved has been added as an illustrative variable, on order to better link our findings to the descriptions realized by \cite{Balestraetal16,Arazyetal17}. In terms of methodology, it is noticeable to remark that simple data reduction techniques such as clustering and PCA allow to reach a comparable level of information as more refined approaches, such as \cite{Weietal15}, who applied Non-parametric Hidden Markov Clustering models of profiles.

\subsection{Limitations and future research}
However, this work suffers from some limitations that should be discussed, while opening future research direction. First, a strong hypothesis has been made by focusing only on contributors with more than 100 edits. If a potential application of such a clustering approach is to increase the users retention rate, it would be relevant to pay a special attention to the small contributors with less than 100 edits, and design retention strategies for them. However, dealing with such a population would lead to more data quality issues and uncertainty. A deep analysis of cluster revealed also the presence of peripheral participation periods, but mainly for the people on a mission (Cluster 1) so the first edits are of paramount importance and may need special treatments to distinguish between those learners and the casual contributors (Cluster 2), for instance.
The second limitation relies on the volume of data analyzed: the results should be generalized to bigger datasets like the English, or the French Wikipedia. Nevertheless, our research methodology gives some guarantees about the work's generalization capabilities since the methods have been first calibrated on the Romanian Wikipedia and then validated with the Danish Wikipedia (with very good consistency). However, those two are occidental Wikipedias, and it will be as interesting to run the same analysis on Arabic, or Thai, or Hindi wikipedias, in a word, on any other non-occidental, medium size Wikipedia.
Another weakness is related to the limited amount of features used to detect the profiles. It will be relevant to consider other characteristics that will add variety. For instance, a first step would consist in using the number of different articles as an explanatory variable instead of just an illustrative one. Other variables should be added as well, as long as they remain simple and easily observable (and computable) by the project 'managers' in all Wikipedias. 
The highlighted profiles are identifiable early in the history of involvement, suggesting that light monitoring of newcomers may be sufficient to adapt the interaction with them and increase the retention rate. 

But above all, further research will deal with the extension of this offline clustering towards dynamic techniques. The principle is to dynamically adapt the clusters as new contributors join the community. Online clustering methods (such as Growing Neural Gas) could be adapted in order to develop a dynamic decision support tool for online contributors assistance.




\bibliographystyle{SIGCHI}
\bibliography{NJ_KR}

\section*{Appendix. Box Plot Figures}


\begin{figure*}[!htb]
\begin{center}

\includegraphics[width=\textwidth]{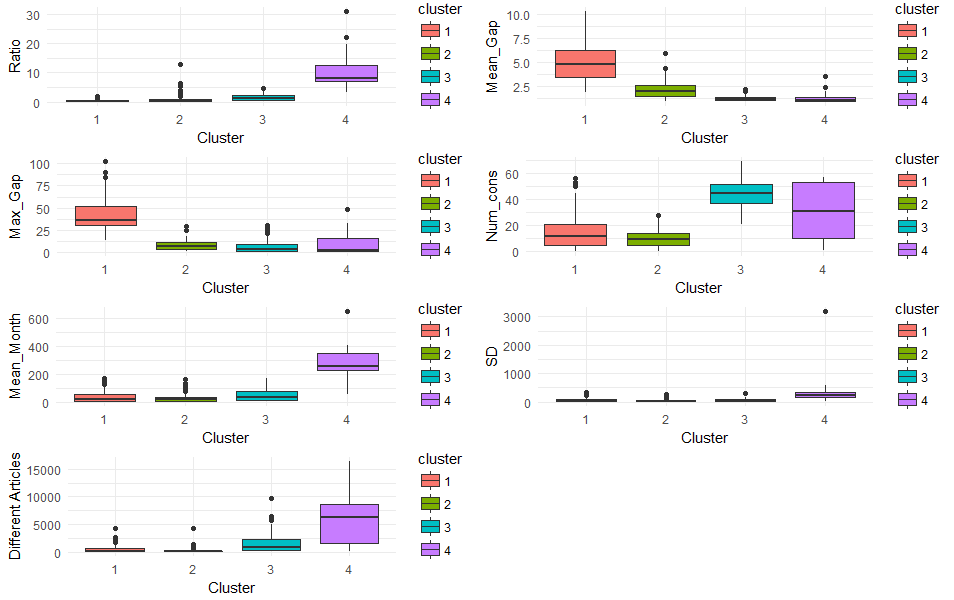}

\caption{Boxplot of features distribution within each cluster for the Danish Wikipedia}
\label{dawiki_boxplots}

\includegraphics[width=\textwidth]{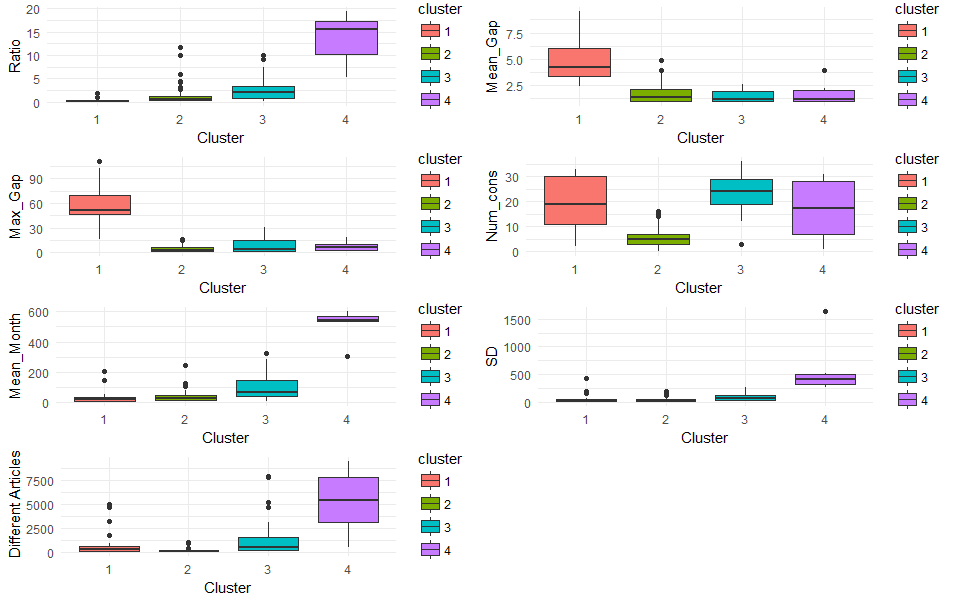}
\caption{Boxplot of features distribution within each cluster for the Romanian Wikipedia}
\label{rowiki_boxplots}
\end{center}
\end{figure*}

\end{document}